# Elemental topological Dirac semimetal α-Sn with high quantum mobility

Le Duc Anh,[1,2,3,+,*] Kengo Takase,[1,+] Takahiro Chiba[4], Yohei Kota[4], Kosuke Takiguchi,[1]

and Masaaki Tanaka[1,5,**]

[1]*Department of Electrical Engineering and Information Systems, The University of Tokyo, 7-3-1 Hongo, Bunkyo-ku, Tokyo 113-8656, Japan*

[2]*Institute of Engineering Innovation, The University of Tokyo, 7-3-1 Hongo, Bunkyo-ku, Tokyo 113-8656, Japan*

[3]*PRESTO, Japan Science and Technology Agency, 4-1-8 Honcho, Kawaguchi, Saitama, 332-0012, Japan*

[4]*National Institute of Technology, Fukushima College, Iwaki, Fukushima, 970-8034, Japan*

[5]*Center for Spintronics Research Network (CSRN), The University of Tokyo, 7-3-1 Hongo, Bunkyo-ku, Tokyo 113-8656, Japan*

E-mail: *\*anh@cryst.t.u-tokyo.ac.jp*,
*\*\*masaaki@ee.t.u-tokyo.ac.jp*
+ These authors contributed equally

## Abstract

α-Sn with a diamond-type crystal structure provides an ideal avenue to investigate novel topological properties owing to its rich diagram of topological phases and simple elemental material structure. Thus far, however, realisation of high-quality α-Sn remains a challenge, which limits our understanding of its quantum transport properties and device applications. Here, we present epitaxial growth of α-Sn on InSb (001) with the highest quality thus far and reveal that it is a topological Dirac semimetal (TDS) by quantum transport investigations together with first-principles calculations. We realise unprecedentedly high quantum mobilities of both the surface state (30000 $cm^2$/Vs), which is ten times higher than the previously reported values, and the bulk heavy-hole (HH) state (1700 cm$^2$/Vs), which has never been obtained experimentally. These excellent features allow us, for the first time, to quantitatively characterise the nontrivial interfacial and bulk band structure of α-Sn via Shubnikov-de Haas oscillations at various temperatures and under various magnetic field directions. These results reveal the existence of a topological surface state (TSS) and a bulk HH band, both with nontrivial phase shifts, indicating that the TDS phase of α-Sn is established. Furthermore, we demonstrate a crossover from the TDS to a two-dimensional topological insulator (2D-TI) and a subsequent phase transition to a trivial insulator when varying the thickness of α-Sn. Our work indicates that α-Sn is an excellent model system to study novel topological phases and a prominent material candidate for topological devices.

**Main**

Application of the mathematical concept of topology to the classification of band structures in solid-state physics has led to remarkable discoveries of many exotic topological materials[1,2]. Thus far, topological insulators (TIs)[1,3], Dirac semimetals[2,4,5] and Weyl semimetals[2,6] are widely known, all of which host linear dispersion bands (Dirac cones) protected by their nontrivial topological invariant. These high-mobility massless carriers with robust spin-momentum locking have provided great motivation from application perspectives, such as inducing a dissipationless spin current[1,7], enabling perfect spin-charge conversion in various spintronic devices and inducing exotic emergent states such as Majorana fermions[8], which are essential for topological quantum computing. Among all material candidates, topological Dirac semimetals (TDSs), which contain three-dimensional (3D) Dirac cones in their bulk band structure[4], are of fundamental importance as a parent state of various other topological phases[2,9,10,11,12,13,14].

For example, phase transitions from a TDS to a weak TI or a topological crystalline insulator by varying physical parameters are expected[11]. The quantum size effect in ultrathin TDS films leads to a crossover to a two-dimensional TI (2D-TI)[12,13], where important transport phenomena such as the quantum spin Hall effect[15] and quantum anomalous Hall effect[16] can be expected. Furthermore, breaking time reversal symmetry (TRS) by applying an external magnetic field or magnetic exchange coupling lifts the Kramers degeneracy in the bulk Dirac cones and transforms a TDS into a magnetic Weyl semimetal[2,10]. A TDS itself also exhibits many attractive features, such as ultrahigh mobility[17,18], light cyclotron mass[19], giant linear longitudinal magnetoresistance[17,20], Fermi arcs[11,12,21,22], and the chiral anomaly[23,24,25,26]. However, TDS materials have rarely been realised thus far[11]; the only experimentally identified cases are Na$_3$Bi[4,21,22] and Cd$_3$As$_2$[5,12,13,17,18,19,20], except for one report on α-Sn (111)[14].

Among many topological materials, α-Sn[30,32,34,35,36,37,38] stands out as a unique and promising candidate: It is the only elemental material that shows multiple topological phases, which can be controlled by various means such as applying strain, varying the thickness, incorporating magnetism and applying electric fields[9,10]. Bulk α-Sn is a so-called zero-gap semiconductor, where there is an inverted band order[40] between the *s*-derived $\varGamma_7^-$ band and *p*-derived $\varGamma_8^+$ bands, which consist of an inverted light hole (iLH) band and a heavy-hole (HH) band, as indicated by the calculated band structure shown in the centre panel of Fig. 1a (see Methods). This inverted band order hosts a topological surface state (TSS) that is superimposed by the HH band[32,33]. Under an in-plane compressive strain, the HH band (iLH band) shifts up (down), which results in the emergence of two stable 3D Dirac cones at the band crossing point $\pm \boldsymbol{k}_D = (0,0, \pm k_D)$ protected by a 4-fold rotational symmetry[9,10], as shown in the right panel of Fig. 1a. Since the $Z_2$ invariant $v_{2D}$ defined on the $k_z = 0$ plane is unity, the compressively strained α-Sn (001) has long been expected to be a TDS[3,11]. Furthermore, a phase transition from a TDS to a 2D-TI can be achieved by quantum confinement, i.e., by decreasing the thickness of α-Sn[12,13], as will be discussed later in this work. In contrast, an in-plane tensile strain opens a gap at the $\Gamma$ point and transforms α-Sn into a strong 3D-TI[9,10,14] (left panel of Fig. 1a). Its large topological gap (~400 meV)[29] and high Fermi velocity (~$7.3 \times 10^5$ m/s)[30] provide robustness against thermal fluctuations and a high-speed response, respectively, which are crucial for practical spintronic applications[31]. All of these rich topological phases can be realised on a simple elemental material platform that is compatible with conventional III-V semiconductors such as InSb. Thus, α-Sn is immune to material control problems such as disorder or deviation from stoichiometry, unlike other compound topological materials[27,28]. These important features inarguably position α-Sn as one of the most promising platforms for systematic studies of various topological physics and devices.

However, thus far, only a few studies have accurately captured the nontrivial features of α-Sn, especially its quantum transport properties[34,35,36,37,38]. The main reason for this lack of study is the insufficient crystal quality because α-Sn can easily transform into the more stable trivial β-Sn phase or because the nontrivial features of α-Sn are impaired by the interface roughness when grown on substrates. In previous studies, α-Sn was grown directly on an ion-milled InSb substrate surface, which deteriorates the interface and bulk quality[30,32,33,34,36,37,39]. In another work, Bi or Te was deposited as a surfactant during α-Sn growth, which unwantedly voided the advantage of the elemental topological material α-Sn[39]. High-quality and well-controlled α-Sn is thus urgently required to realise its full potential.

In this work, we demonstrate epitaxial growth of very high-quality α-Sn thin films on InSb (001) substrates using molecular beam epitaxy (MBE). The unprecedentedly high crystal quality and mobility allow us, for the first time, to accurately characterise the nontrivial bulk and surface band structures of α-Sn, and we show that these α-Sn thin films are indeed in the quantised TDS phase by quantum transport measurements.

First, an α-Sn thin film (9.2 nm, 56 monolayers (MLs)) is grown on an InSb (001) substrate at low temperature (~ –5 °C) using MBE (see Methods and Supplementary Fig. S1). Prior to the growth of α-Sn, a 200 nm-thick InSb buffer layer is grown to obtain an atomically flat interface, which is terminated by an In-stabilised topmost surface. A cross-sectional scanning transmission electron microscopy (STEM) lattice image clearly indicates a high-quality diamond-type crystal structure of α-Sn and a perfectly flat interface with the InSb buffer layer, as shown in Fig. 1c. The transmission electron diffraction (TED) pattern of the α-Sn layer also confirms a diamond structure without any other precipitates, such as tetragonal β-Sn (Fig. 1d). Moreover, the average roughness estimated by atomic force microscopy (AFM) is 0.17 nm, indicating an atomically flat α-Sn surface (Fig. 1e). From the TED data, we estimate the in-plane and perpendicular-to-the-plane lattice constants of the α-Sn thin film, which indicate that a strong in-plane epitaxial compressive strain of –0.76% is applied in the α-Sn layer (see Methods). This strong compressive strain is expected to drive the system into a TDS phase, as indicated by our first-principles calculations shown in Fig. 1b.

We first characterise the 56 ML-thick α-Sn sample using magneto-transport measurements (see Methods). As shown in Fig. 2a, the longitudinal resistance $R_{xx}$ exhibits strong and clear Shubnikov-de Haas (SdH) oscillations, which appear at a magnetic field $B$ as small as 0.3 T at 2 K (see also Supplementary Fig. S2) and persist up to 20 K, manifesting the high sample quality. The first derivative of the longitudinal conductance $dG_{xx}/dB$ (Fig. 2b) and Fourier transformations of the oscillatory part $\Delta G_{xx}$ (Figs. 2c,f, see Methods) reveal three components in the SdH oscillations: Two oscillations with close frequencies ($F_1$ = 13.4 and $F_2$ = 15.1 T) form a beat pattern at a low magnetic field ($B^{-1}$ > 0.45 T$^{-1}$), and the other oscillation with a relatively high frequency ($F_3$ = 37.2 T) starts at a high magnetic field ($B^{-1}$ < 0.2 T$^{-1}$). The $i$ component of $\Delta G_{xx}$ can be fitted based on the Lifshitz-Kosevich (LK) theory, expressed by [41]

$$\Delta G_i(B, T, F_i, m_i, \tau_{Di}, \gamma_i)$$
$$= G_{0i} \frac{2\pi^2 k_B T/\hbar\omega_{ci}}{\sinh(2\pi^2 k_B T/\hbar\omega_{ci})} \exp\left(-\frac{\pi}{\omega_{ci}\tau_{Di}}\right) \cos\left(2\pi\left(\frac{F_i}{B} - \gamma_i\right)\right) \quad (1)$$

Here, $G_{0i}$ is the proportionality coefficient, $\hbar$ is the Dirac constant, $k_B$ is the Boltzmann constant, $\tau_{Di}$ is a quantum relaxation time, $m_i$ is the cyclotron mass, $\omega_{ci} = qB/m_i$ is the cyclotron angular frequency, where $q$ is the elementary charge, and $\gamma_i$ is a phase shift determined by the band topology. Here, we use LK theory for the case of 2D transport[41] because all three band components are quantised in the growth direction, as shown later. Note that $\gamma_i = 0, 1$ is expected for massless Dirac fermions with a linear dispersion, whereas $\gamma_i = 1/2$ is expected for massive fermions in trivial band components. By fitting this expression to the temperature dependence of the Fourier-transformed (FT) intensities (see Methods), we estimate the three cyclotron masses to be $m_1$ = 0.035 $m_0$, $m_2$ = 0.035 $m_0$, and $m_3$ = 0.11 $m_0$, as shown in the insets of Figs. 2c,f. Using these mass values, we fit eq. (1) to the $\Delta G_{xx}$ measured at $T$ = 2 K in the low magnetic field range ($B^{-1}$ > 0.45 T$^{-1}$), where only components $F_1$ and $F_2$ are superimposed, by taking a sum of the two oscillations expressed by eq. (1) with $G_{01}, \tau_{D1}, \gamma_1$, $G_{02}, \tau_{D2},$ and $\gamma_2$ as the fitting parameters (Fig. 2d and Supplementary Fig. S4). The following values are obtained: $\tau_{D1}$

= 495 ± 11 fs and $\gamma_1$ = 0.5325 ± 0.0009 for $F_1$ and $\tau_{D2}$ = 611 ± 20 fs and $\gamma_2$ = 0.8098 ± 0.0012 for $F_2$ (here, error bars are the standard errors of the fitting process). The quantum mobilities $\mu_{Di} = q\tau_{Di}/m_i$ are then estimated to be $\mu_{D1}$ = 24900 ± 600 cm²/Vs and $\mu_{D2}$ = 30000 ± 1000 cm²/Vs for the two components $F_1$ and $F_2$, respectively. Additionally, we fit eq. (1) to $\Delta G_{xx}$ in a high magnetic field range ($B^{-1}$ < 0.2 T$^{-1}$), where component $F_3$ dominates, as shown in Figs. 2f,g. Here, to enhance the fitting accuracy, we separately estimate $\tau_{D3}$ = 109 ± 33 fs and $\gamma_3$ = 0.71 ± 0.06 using a Dingle plot and a fan plot (see Methods and Supplementary Fig. S5). This reveals a relatively high quantum mobility $\mu_{D3}$ = 1700 ± 500 cm²/Vs for this heavy carrier component. We note that these estimated quantum mobilities are consistent with the critical magnetic field $B_{ci}$ where the corresponding SdH oscillations begin (i.e., they satisfy $\mu_{Di}B_{ci}\sim 1$), which supports the validity of our analysis of the SdH oscillations.

By combining them with first-principles calculations, these estimated parameters reveal essential details of the band structure of 56 ML-thick α-Sn. Our calculated band structure in Fig. 1b (see also Supplementary Figs. S7 and S8) clearly shows a linear dispersion in the surface states (green data points) with a Dirac point (DP) at $E_{DP}$, derived from the topmost Sn ML, which is the TSS, and a quantised bulk band (blue points), which opens a small gap of 30 meV (yellow square) at the bulk Dirac point at zero energy. Among the three components derived from quantum transport, remarkably, oscillation $F_2$ has a nontrivial phase shift $\gamma_2$ = 0.8098, a light cyclotron mass $m_2$ = 0.035 $m_0$ and a very high quantum mobility of 30000 cm²/Vs. We thus attribute this component to the TSS of α-Sn. The phase shift is slightly off from the ideal unity value, which may be caused by hybridisation of the top and bottom surface states[42] (see Supplementary Note 1). The quantum mobility of the TSS is experimentally estimated for the first time and is one order of magnitude higher than the mobility (~ 3000 cm$^2$/Vs) estimated by the Hall measurements (namely, Hall mobility) reported in all previous works[34,37,38]. The results described above have been achieved because of the high crystal quality and atomically flat interface of our α-Sn/InSb heterostructure (note that for the same sample, the quantum mobility is usually much smaller than the Hall mobility). Meanwhile, the $F_3$ component with a relatively heavy cyclotron mass ($m_3$ = 0.11 $m_0$) can be attributed to the HH band of α-Sn because there are no other carriers in α-Sn and InSb that have such a heavy cyclotron mass[43]. The high quantum mobility of 1700 cm$^2$/Vs of the HH band allows us, for the first time, to elucidate the detailed properties of the bulk Dirac cones in α-Sn via the SdH oscillations. The nontrivial phase shift ($\gamma_3$ = ~0.71) of the HH band indicates that the TDS phase is achieved as expected. This phase shift of the HH band deviates from the ideal phase shift of unity, possibly because of the gap opening due to the quantum confinement effect[42]. We note that this phase shift of the HH band does not originate from hybridisation with the TSS because such hybridisation is in principle prohibited by their different wavefunction symmetries[33,44]. Furthermore, from the SdH frequency $F_2$, we can estimate the Fermi level $E_F$ with respect to the $E_{DP}$ of the TSS, given as

$$|E_{\mathrm{DP}} - E_F| = \frac{q\hbar F_2}{m_2} = 50 \text{ meV} \qquad (2)$$

According to this value, the position of $E_F$ is depicted as a dotted pink line in Fig. 1b, which crosses both the upper linear dispersion band of the TSS (see also Supplementary Figs. S7 and S8) and the HH band top. Additionally, we attribute component $F_1$, which has a trivial phase shift and a light cyclotron mass, to the conduction band of InSb, where parallel conduction occurs (see Supplementary Note 2).

Our calculated result in Fig. 1b indicates that the 56 ML-thick α-Sn is in a quantised TDS phase. The angular dependence of the SdH oscillations, which is shown in Fig. 3, confirms this description of the α-Sn band structure. We rotate the magnetic field $B$ from the perpendicular [001] direction, where $B \perp I$, to the in-plane [$\bar{1}$10] direction, where $B // I$. Here, $\theta$ is defined as the angle of $B$ with respect to the [001] direction. The $B$ dependence of $R_{xx}$ changes from positive to negative when $B$ is rotated from the $B \perp I$ to $B // I$ configuration, as seen in Fig. 3a. The origin of this negative magnetoresistance for $B // I$ may be attributed to the chiral anomaly, a characteristic feature of the TDS phase[9,23,24,25,26]. Furthermore, the frequencies of the TSS ($F_2$) and HH ($F_3$) bands vary as $1/\cos\theta$ over a wide range of $\theta$ up to 60° (Fig. 3b), which confirms the 2D nature of these bands. The $F_2$ component is the TSS, which is of course a 2D band. Interestingly, the HH band ($F_3$) also shows 2D behaviour in our sample, which is consistent with the long coherence length $l_3$ ($= v_F\tau = 40$ nm) compared with the thickness of the α-Sn layer (9.2 nm). This is reasonable considering the band offset of ~0.5 eV between the $\Gamma_8^+$ bands of Sn and InSb[30,45,46], which leads to the quantum size effect in the HH and iLH bands.

Furthermore, with decreasing α-Sn thickness, we expect a crossover from a TDS to a 2D-TI and eventually a phase transition to a normal insulator (NI) due to the strong quantum confinement[12,13]. This is clearly indicated in our first-principles calculations at different α-Sn thicknesses, which are shown in Fig. 4a. The energy gap at the $\Gamma$ point of α-Sn, shown in the bottom panel of Fig. 4b, first increases with decreasing thickness from 100 MLs, which indicates the crossover between the quantised TDS and 2D-TI phases in this region (40 ~ 100 MLs). Then, when the α-Sn thickness is decreased to 20 ~ 28 MLs, we observe a quick decrease, closing, and reopening of the energy gap, reflecting the topological phase transition from a nontrivial (2D-TI) to a trivial insulator (NI), associated with a change in the $Z_2$ invariant (top panel of Fig. 4b). Note that the critical thickness (20 ~ 28 MLs) is thicker than that (10 MLs) of previous studies[47], which seems to be caused by the strong compressive strain in our α-Sn thin films. We clearly observe such a phase transition in the SdH oscillations measured in a series of α-Sn samples with different thicknesses (see Supplementary Fig. S9). The development of the SdH frequencies of all three components TSS ($F_2$), HH ($F_3$), and InSb ($F_1$) is shown in Fig. 4c, which clearly indicates that the TSS component disappears with decreasing thickness of α-Sn below a critical thickness between 30 and 40 MLs. This behaviour is consistent with the calculated band structures in Fig. 4a and Supplementary Fig. S7. This may be caused by the surface hybridisation of the top and bottom surface states and the strong compressive strain (see Supplementary Note 1). The decrease in $F_2$ and $F_3$ may also be related to a change in the $E_F$ position with deceasing thickness. Notably, the phase shift of the HH band is always nontrivial within the error bar (see top panel of Fig. 4c) regardless of whether the TSS exists, which supports the conclusion that the nontrivial phase shift of the HH band originates from the bulk DP rather than from hybridisation with the TSS. These results demonstrate that α-Sn is indeed an elemental TDS, which can be a fundamental platform for investigating various novel topological phases.

In summary, an elemental TDS phase is confirmed in α-Sn (001) under in-plane compressive strain for the first time by our quantum transport studies and first-principles calculations. The high-quality α-Sn thin film with very high quantum mobilities of the TSS (30000 cm$^2$/Vs) and HH state (1700 cm$^2$/Vs) enables an unprecedentedly accurate description of the nontrivial band structure of α-Sn. A crossover from a TDS to a 2D-TI and a subsequent phase transition to a trivial insulator are also demonstrated by varying the thickness of α-Sn. Our results prove that high-quality crystal growth of α-Sn is the key to unlocking its rich potential for topological physics and device applications.

**Methods**

**Sample preparation and characterisation**

A α-Sn thin film (9.2 nm, 56 monolayers (MLs)) is grown *in situ* on an InSb (001) substrate without any surfactant using our molecular beam epitaxy (MBE) system equipped with effusion cells of Sn and III-V elements (Supplementary Fig. S1a). Prior to the growth of α-Sn, a 200 nm-thick InSb buffer layer is grown to obtain an atomically flat α-Sn/InSb interface. *In situ* reflection high-energy electron diffraction (RHEED) along the [110] axis shows streaky patterns (Supplementary Figs. S1b,c), indicating a 2D growth mode of both α-Sn with a diamond-type crystal structure and InSb with a zinc-blende-type crystal structure throughout the MBE growth. An In-stabilised InSb topmost surface that shows c(8×2) reconstruction is prepared, which is similar to previous works in which α-Sn is grown directly on an InSb substrate[30,32,33,34,36,37,39] (Supplementary Fig. S1b). The cross-sectional scanning transmission electron microscopy (STEM) lattice image (Fig. 1c) clearly indicates a high-quality epitaxial α-Sn thin film of a diamond crystal structure with no visible defects and an atomically flat α-Sn/InSb interface. The transmission electron diffraction (TED) pattern of the α-Sn layer (Fig. 1d) also confirms that the diamond structure is maintained without any other precipitates, such as β-Sn, which has a tetragonal structure. Moreover, the root-mean-square (RMS) roughness obtained from the atomic force microscopy (AFM) image (Fig. 1e) is 0.171 nm, confirming the very flat surface. We note that strong in-plane compressive strain is applied on the present α-Sn layer compared with that in previous reports of α-Sn/InSb (001). This compressive strain drives the system into the topological Dirac semimetal (TDS) phase.

To prepare α-Sn films with different thicknesses, the 56 ML-thick sample is gradually etched by Ar-ion milling to 39, 30, and 0 MLs. The etching process is conducted in combination with secondary ion mass spectrometry (SIMS), which allows us to accurately control the etching depth. A 21 ML-thick α-Sn thin film is separately grown on InSb (001). The 0 ML-thick sample serves as a reference sample to study the parallel conduction in the InSb buffer layer and the substrate.

**Estimation of the in-plane compressive strain in α-Sn**

From the TED pattern (Fig. 1d), we obtain lattice spacings $d_{002}$ = 3.33 Å and $d_{-111}$ = 3.76 Å. The in-plane and out-of-plane lattice constants $a_{//}$ and $a_\perp$ and the in-plane strain $\epsilon_{//}$ of α-Sn are estimated as follows.

$$a_\perp = 2 \times d_{002} = 6.66 \text{ Å} \qquad (4)$$

$$a_{//} = \sqrt{\frac{2}{(d_{002}^{-1})^2 + (d_{-111}^{-1})^2}} = 6.44 \text{ Å} \qquad (5)$$

$$\epsilon_{//} = \frac{a_{//} - a_i}{a_i} = -0.76\% \qquad (6)$$

Here, $a_i$ = 6.4892 Å is the intrinsic lattice constant of the α-Sn bulk[48]. The in-plane compressive strain in our sample ($\epsilon_{//} = -0.76\%$) is larger than that ($\epsilon_{//} = -0.14\%$) in the α-Sn/InSb (001) heterostructure reported previously[49]. This may be caused by some residual surface oxides at the InSb buffer/InSb substrate interface. We note that such precipitates only affect the strain of the system and do not void the high crystal quality of α-Sn, which is clearly confirmed by the STEM lattice image (Fig. 1c).

**Magneto-transport measurements**

Hall bar devices with sizes of approximately $500 \times 2000$ um$^2$ are formed manually by scratching the sample surface to avoid any unwanted heating during the standard lithography process. Magneto-transport measurements are performed at various temperatures (2 ~ 20 K) using a standard four-point method with a constant current (100 µA) applied in the $[\bar{1}10]$ direction and a magnetic field up to 14 T. Each measurement gives a longitudinal resistance $R_{xx}$ and a Hall resistance $R_{yx}$, from which the conductance $G_{xx} = R_{xx}/(R_{xx}^2+R_{yx}^2)$ is estimated, as shown in Supplementary Fig. S2. The magnetic field direction dependence of $R_{xx}$ is measured by rotating the sample in a fixed magnetic field from the perpendicular [001] direction ($\boldsymbol{B}\perp\boldsymbol{I}$) to the in-plane $[\bar{1}10]$ direction ($\boldsymbol{B}//\boldsymbol{I}$), as shown in Figs. 3a,b. $\theta$ is defined as the angle between the magnetic field $\boldsymbol{B}$ and the [001] direction.

**Polynomial fitting procedure to extract the oscillatory part of $G_{xx}$**

To analyse the Shubnikov-de Haas (SdH) oscillations by Lifshitz–Kosevich (LK) theory, the oscillatory part $\Delta G_{xx}$ is extracted from the raw experimental data $G_{xx}$ after removing the background component of $G_{xx}$ using polynomial fitting (Supplementary Fig. S3). Before the polynomial fitting, the experimental data of $G_{xx}$ at a high magnetic field range (0.071 T$^{-1}$ < $B^{-1}$ < 0.2 T$^{-1}$) are interpolated with a constant interval of $B^{-1}$ = 0.001 T$^{-1}$. The purpose of this is to avoid artificial weighting of the polynomial fitting of $G_{xx}(B^{-1})$ at low $B$ values because the raw data are obtained with a constant interval of $B$, which causes large distances between two consecutive data points along the $B^{-1}$ axis in the low $B$ range. Note that this procedure does not degrade the validity of the analysis. For the $G_{xx}$ at a low magnetic field range ($B^{-1}$ > 0.45 T$^{-1}$), we conduct detailed measurements with small magnetic field steps so that the interpolation process is not necessary. To improve the fitting accuracy, we use different polynomials for the data in different ranges of magnetic field: A polynomial of degree 11 provides the best fit in the range of 0.45 T$^{-1}$ < $B^{-1}$ < 2.5 T$^{-1}$ (Supplementary Fig. S3a), where there are only two components $F_1$ and $F_2$, while a cubic polynomial is used for fitting to $G_{xx}$ in the range of 0.071 T$^{-1}$ < $B^{-1}$ < 0.2 T$^{-1}$ (Supplementary Fig. S3b), where the component $F_3$ dominates. In both cases, the polynomials well reproduce the background components of $G_{xx}$. The $G_{xx}$ data in the very low-magnetic-field range ($B^{-1}$ > 2.5 T$^{-1}$), where there is no SdH oscillation, and in the intermediate range (0.2 T$^{-1}$ < $B^{-1}$ < 0.45 T$^{-1}$), where all the three components $F_1$, $F_2$, $F_3$ coexist and complicate the situation, were not taken into our polynomial fitting and further analysis. The same procedure is applied to obtain the oscillatory part $\Delta G_{xx}$ in all the $G_{xx}$ measured at different temperatures and thicknesses (data not shown).

**Analysis of SdH oscillations**

The temperature dependence of the Fourier-transformed peak intensity of each component of the SdH oscillations is fitted by the temperature-dependent part of the LK theory expression, given in eq. (1), as shown below[41] (insets of Figs. 2c,f).

$$G_T(T; B, m_i) \propto \frac{\frac{2\pi^2 k_B T}{\hbar \omega_c}}{\sinh\left(\frac{2\pi^2 k_B T}{\hbar \omega_c}\right)} \quad (7)$$

In this equation, the magnetic field $B$ is defined as $B^{-1}$ = ($B_{max}^{-1}$+$B_{min}^{-1}$)/2, where $B_{max}$ ($B_{min}$) is the maximum (minimum) value of the magnetic field range used in the fitting. This fitting gives the cyclotron masses $m_1$, $m_2$, and $m_3$ of the three components $F_1$, $F_2$ and $F_3$, respectively.

For the fittings in Figs. 2d,g, we first deduce the oscillatory part $\Delta G_{xx}$ as mentioned in the previous subsection. At the low magnetic field range ($B^{-1}$ > 0.45 T$^{-1}$) where $F_1$ and $F_2$ are superimposed, the $\Delta G_{xx}$ at $T$ = 2 K is fitted by a sum of two oscillations described by eq. (1) based on the standard LK theory, as shown in Fig. 2d and Supplementary Fig. S4. $G_{01}, \tau_{D1}, \gamma_1, G_{02}, \tau_{D2},$ and $\gamma_2$ are the fitting parameters, while other parameters $F_1$, $m_1$, $F_2$, and $m_2$ are fixed to improve the accuracy of the fitting. Additionally, at high magnetic fields ($B^{-1} < 0.2$ T$^{-1}$), where the quickly oscillating component $F_3$ dominates, $\tau_{D3}$ and $\gamma_3$ are independently determined by a standard Dingle plot and a fan plot, shown in Supplementary Figs. S5, using eqs. (8) and (9), respectively.

$$\ln \frac{\Delta G_{xx}}{\frac{2\pi^2 k_B T \hbar \omega_c}{\sinh 2\pi^2 k_B T \hbar \omega_c}} = -\frac{m_3 \pi}{q\tau_{D3}} \frac{1}{B} + C \qquad (8)$$

$$2\pi \left(\frac{F}{B} - \gamma\right) = (2N - 1)\pi \qquad (9)$$

We note that, as shown in Supplementary Fig. S5b, the fitting results of the phase shift are not greatly affected even if we assume that the phase shift depends on $B$, as recently suggested by Wright and McKenzie[42] (WM), given by eq. (10),

$$2\pi \left(\frac{F}{B} + A_1 + A_2 B\right) = (2N - 1)\pi \qquad (10)$$

Here, the $B$-independent term $-A_1$ corresponds to the phase shift $\gamma_3$. The $\gamma_3$ value of 0.71 estimated by LK theory is very close to the $-A_1$ value of 0.73 estimated by WM theory; this good agreement supports the reliability of the estimated phase shift. Eventually, $\Delta G_{xx}$ at $T = 2$ K at a high magnetic field is fitted based on LK theory (eq. (1)), as shown in Fig. 2g, where $G_{03}$ is only the fitting parameter, while the other parameters are fixed.

**First-principles calculations of the band structure of α-Sn**

First-principles calculations based on density functional theory are performed using the Vienna ab initio simulation package (VASP) with the projector augmented wave method[50,51]. The exchange–correlation functional is described under the local density approximation (LDA). Spin–orbit interaction is taken into account in the self-consistent calculations of the electronic structure. We adopt a α-Sn lattice constant of $a = 6.4746$ Å, which is determined by the total energy minimisation in a previous LDA study[52] (this value is in excellent agreement with a measured value, $a$ = 6.4798 Å[53]). We use the tetragonal unit cell including four Sn atoms for the bulk system and the layered unit cell deposited along the [001] direction for the slab geometries. Specifically, slab geometries consisting of 10, 28, 40, 56, and 100 MLs are adopted for the computation of the topological surface state (TSS) of strained α-Sn. Both edges of slabs are terminated with H atoms to remove dangling bonds. Additionally, the surfaces are separated by a vacuum layer with a thickness of 18 Å. The cutoff energy for the plane wave basis is fixed to 500 eV and 300 eV in the bulk and slab geometry calculations, respectively, and the Brillouin zone integrations are replaced by a sum over $15 \times 15 \times 10$ and $8 \times 8 \times 1$ Monkhorst-Pack $k$-point meshes[54]. To simulate the strain effect on the topological electronic states, we consider a biaxial in-plane compressive strain of –0.76%. Structural optimisation is performed under the LDA scheme, whereas we employ the LDA+U method to calculate the band dispersion[56]. The correction energy of the on-site Coulomb repulsion is set to $U$ = –4.5 eV[55], which provides a better description of the topological electronic states in the bulk and slab geometries of α-Sn[32,52,55]. We confirm that the electronic structure around the band gap obtained from the LDA+U approach reproduces the calculation result from the HSE06 hybrid functional in the bulk system[9,10,52].

**Acknowledgements**

This work was partly supported by Grants-in-Aid for Scientific Research (18H05345, 20H05650, 20K15163, 20H02196), the CREST (JPMJCR1777) and PRESTO (JPMJPR19LB) Programs of JST, and the Spintronics Research Network of Japan (Spin-RNJ). A part of this work was conducted at the Advanced Characterization Nanotechnology Platform of the University of Tokyo, supported by the "Nanotechnology Platform" of the Ministry of Education, Culture, Sports, Science and Technology (MEXT), Japan. A part of the computations in this work were carried out using the facilities of the Supercomputer Center, the Institute for Solid State Physics, the University of Tokyo.

**Author contributions**

L. D. A. and K. Takase designed the experiments and grew the samples. K. Takase performed sample characterisations and transport property investigations. K. Takase and K. Takiguchi performed transport data analysis. T. C. and Y. K. performed theoretical calculations. L. D. A. and M. T. planned the study. All the authors intensively discussed and wrote the manuscript.

**Data availability.** The data that support the findings of this study are available from the corresponding authors upon reasonable request.

**Competing financial interests**

The authors declare no competing financial interests.

**Figures and captions**

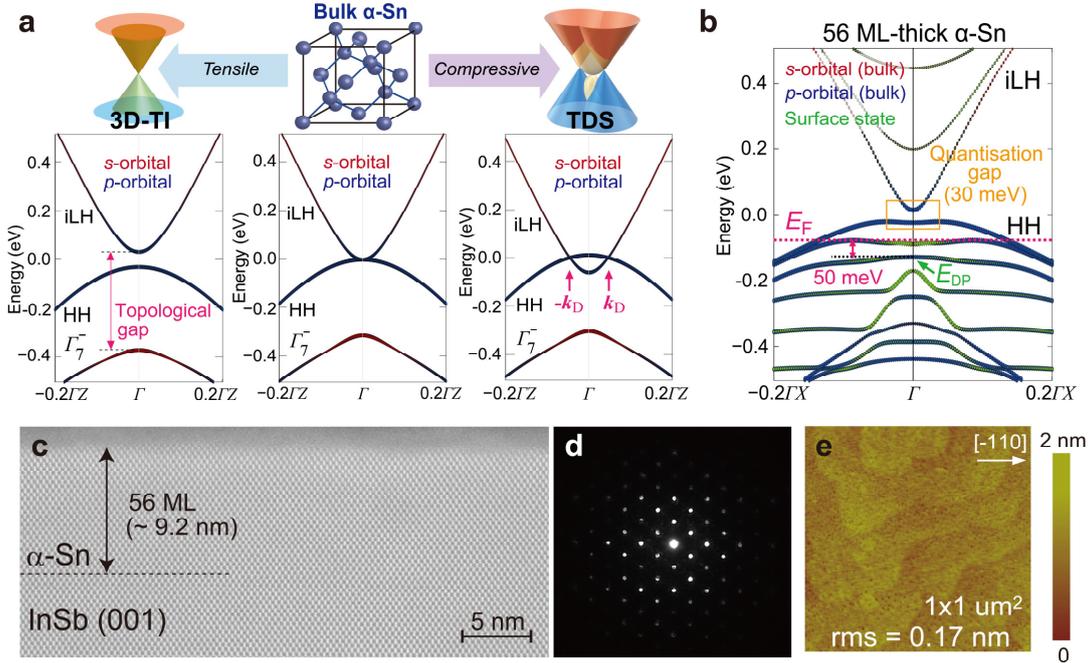

**Fig. 1| Topological phases, band structures, and crystal structure of α-Sn. a.** Bulk α-Sn has a diamond-type crystal structure, whose band structure consists of *p*-derived (blue) $\Gamma_8^+$ bands (iLH and HH) and an *s*-derived (red) $\Gamma_7^-$ band. The origin of the vertical axis corresponds to the Fermi energy of the calculation. The iLH and HH band touching is a 4-fold degeneracy protected by the cubic symmetry. Under an in-plane compressive (tensile) strain, the system is driven into a TDS (3D-TI) phase, as shown in the calculated results in the right (left) panel. Bulk Dirac cones in the TDS phase protected by the rotational symmetry are formed at $\pm \boldsymbol{k}_D = (0,0, \pm k_D)$. **b,** Calculated band structure of 56 ML-thick α-Sn under an in-plane compressive strain of –0.76%. The bulk Dirac cone is slightly gapped (~30 meV) due to the quantum confinement (yellow rectangle). A TSS (green) with a Dirac point at $E_{DP}$, the position of which can be estimated separately from Supplementary Fig. S8, is covered by the HH band. The pink dotted line shows the Fermi level $E_F$ estimated from the analysis of SdH oscillations. **c,d,e.** Cross-sectional STEM lattice image and TED pattern in the [110] direction and top-view AFM image of the 56 ML-thick α-Sn film grown on InSb (001). The root-mean-square (RMS) roughness is 0.17 nm (~1 ML), which indicates an atomically flat surface.

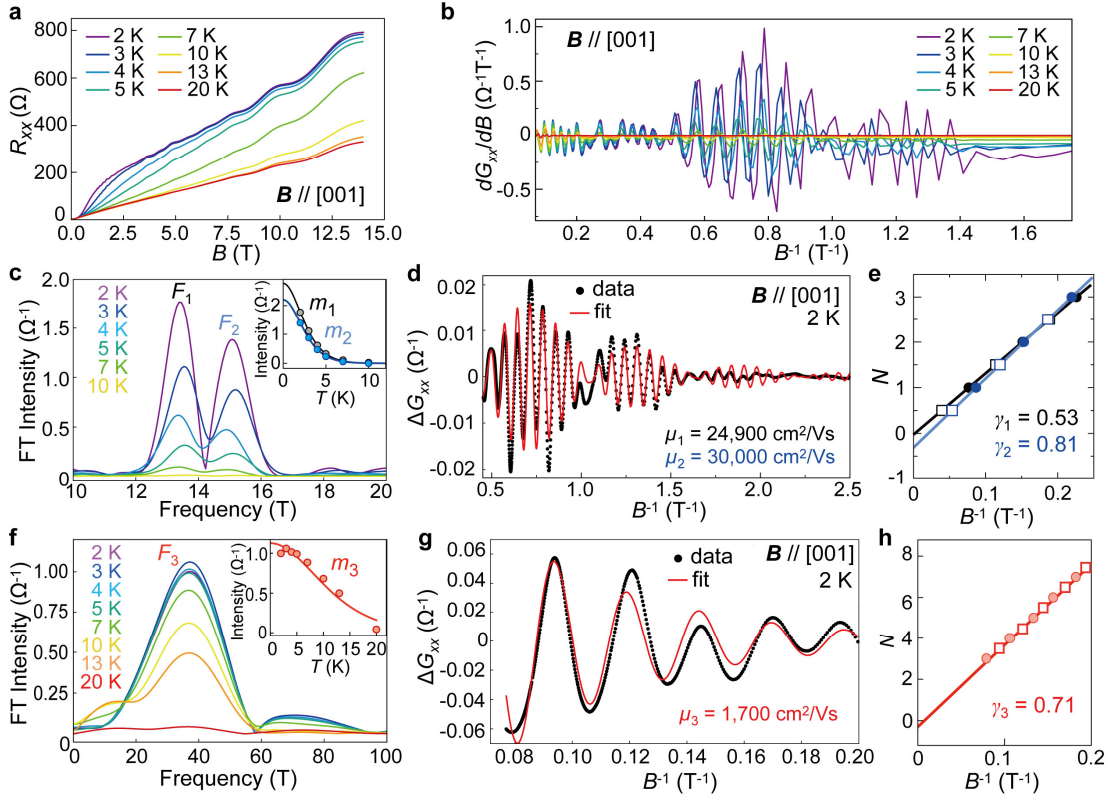

**Fig. 2| SdH oscillations of 56 ML-thick α-Sn under a perpendicular magnetic field. a,** $R_{xx} - B$ under a perpendicular magnetic field $B$ at various temperatures. **b,** First derivative of $G_{xx}$, $dG_{xx}/dB$, as a function of $B^{-1}$ at various temperatures. **c,d,** FT spectra at various temperatures and SdH oscillations at 2 K (black circles) with a fitting curve (red curve) based on two oscillatory components $F_1$ and $F_2$ given by eq. (1) of the oscillatory part $\Delta G_{xx}$ (see Methods) at a low magnetic field ($B^{-1} > 0.45$ T$^{-1}$). **e,** Fan plot of the two components $F_1$ and $F_2$. The maxima (black and blue circles) and minima (black and blue open squares) values are estimated from the fitting curves of the two components (see Supplementary Fig. S4). **f,g,** FT spectra at various temperatures and SdH oscillations at 2 K (black circles) with a fitting curve based on LK theory (red curve) of the oscillatory part $\Delta G_{xx}$ at a high magnetic field ($B^{-1} < 0.2$ T$^{-1}$), where the component $F_3$ dominates. **h,** Fan plot of component $F_3$. The maxima (pink circles) and minima (red open squares) values are estimated from the experimental SdH oscillation shown in **g**. The insets in **c** and **f** show fitting results of the temperature dependence of the FT peak intensities, which give the cyclotron mass values $m_1$, $m_2$, and $m_3$.

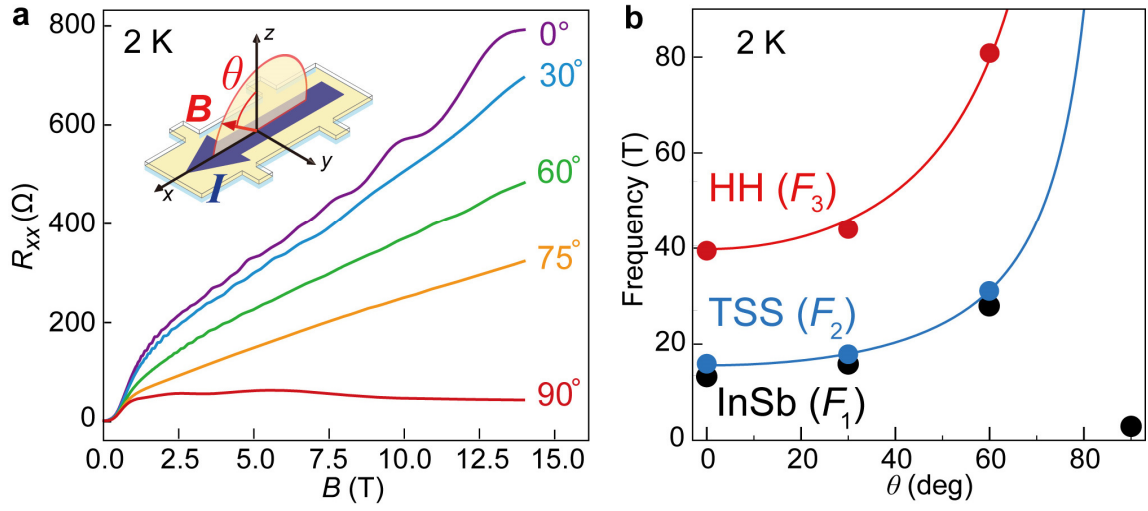

**Fig. 3| Angular dependence of SdH oscillations. a,** Magnetic field direction dependence of the longitudinal resistance $R_{xx}$ vs. $B$ in 56 ML-thick α-Sn. The magnetic field **B** is rotated from the [001] direction (**B**⊥**I**, $\theta = 0°$) to the [$\bar{1}$10] direction (**B**//**I**, $\theta = 90°$). $\theta$ is defined as the rotation angle of **B** with respect to the [001] direction, as shown in the inset. **b,** Angular dependence of the frequencies $F_1$, $F_2$, and $F_3$ (black, blue, and red circles) in the SdH oscillations. Experimental data are fitted by $a/\cos\theta$ (blue and red curves), where $a$ is a proportionality coefficient. These results indicate the 2D characteristics of the TSS ($F_2$) and HH ($F_3$) bands.

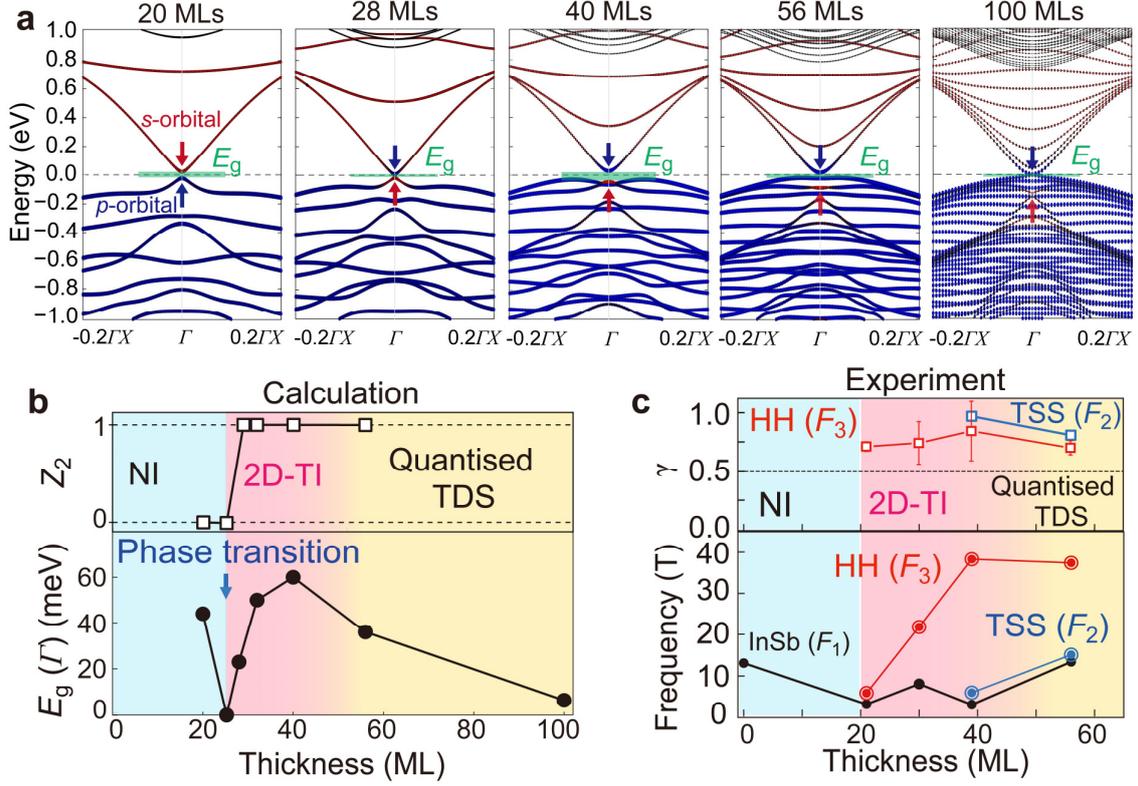

**Fig. 4| Phase transition of α-Sn from a TDS to a 2D-TI and an NI when varying the thickness. a,** Calculated band structure of α-Sn with different thicknesses from 20 to 100 MLs. The origin of the vertical axis corresponds to the Fermi energy of the calculation. Red (blue) points correspond to *s*-(*p*-)derived bands. A band inversion between the *s*- and *p*-derived bands occurs between 20 and 28 MLs, as indicated by the red and blue arrows. Green rectangles mark the energy gap $E_g$ at the $\Gamma$ point, whose magnitudes are summarised in **b**. **b,** (Bottom panel) Calculated thickness dependence of the direct energy gap $E_g$ at the $\Gamma$ point. Corresponding topological phases at various thicknesses are illustrated. (Top panel) Theoretical $Z_2$ invariant (white squares) deduced from the band structure for various thicknesses. $Z_2$ is determined by the parities of all occupied bands at time-reversal invariant momentum ***k*** points, indicating a topological phase transition at 20 - 28 MLs. **c,** Experimental thickness dependence of the oscillation frequencies (bottom panel) and the phase shift values ($\gamma$, top panel) estimated from the SdH oscillations. The TSS disappears below 39 MLs, while the HH band with a nontrivial phase shift persists down to 21 MLs. These results are consistent with the theoretical prediction of a crossover from a quantised TDS (yellow area) to a 2D-TI (pink area) and then a phase transition to an NI (light blue area), as illustrated in the coloured background.

# Supplementary Information

## Elemental topological Dirac semimetal α-Sn with high quantum mobility

**Supplementary Figures**

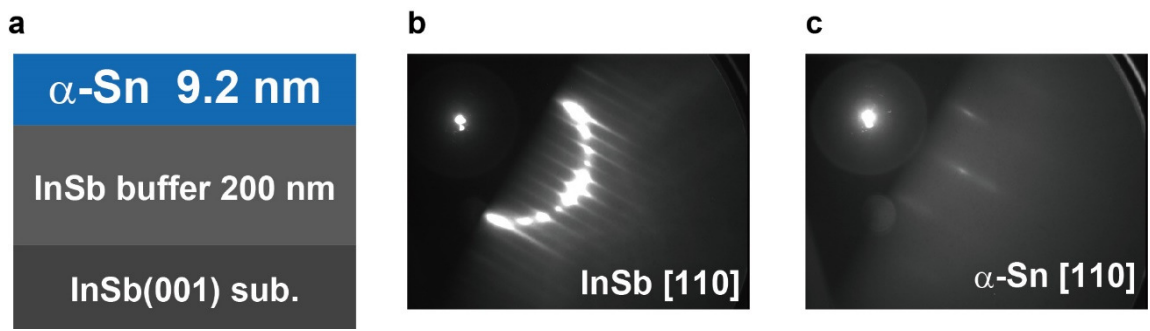

**Supplementary Fig. S1| Sample structure and RHEED patterns during the MBE growth of the 56 ML-thick α-Sn sample. a,** Schematic sample structure. **b, c,** *In situ* reflection high-energy electron diffraction (RHEED) patterns of the InSb buffer and the α-Sn layer, respectively, along the [110] azimuth during the MBE growth.

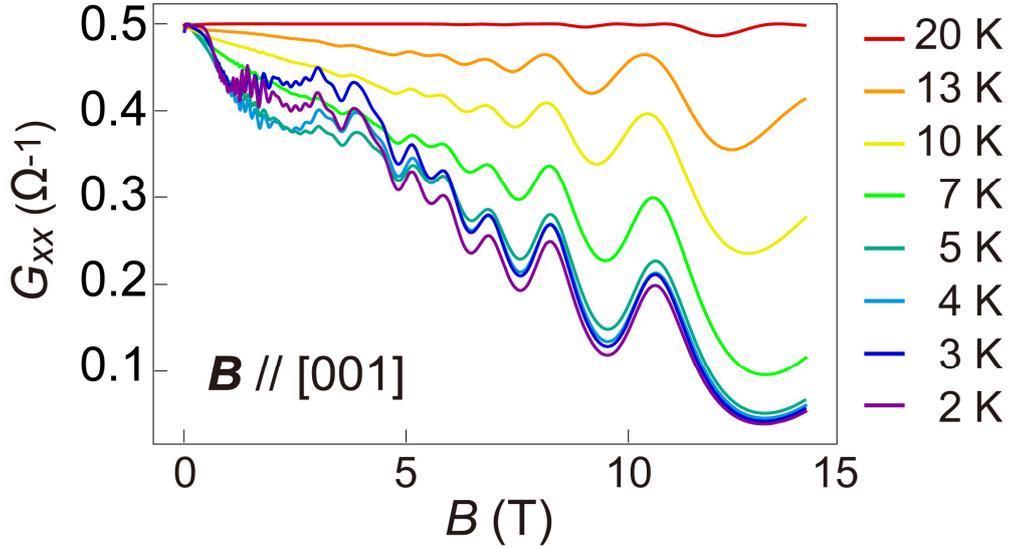

**Supplementary Fig. S2|** $G_{xx}$, estimated from $R_{xx}$ and $R_{yx}$, of the 56 ML-thick α-Sn sample under a perpendicular magnetic field at various temperatures.

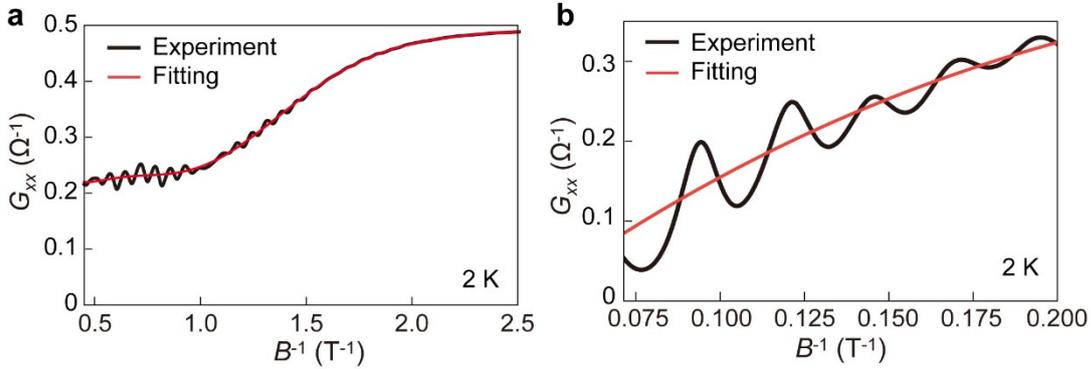

**Supplementary Fig. S3| Polynomial fitting to the background of $G_{xx}$. a,** Experimental result of $G_{xx}$ (black) and the background component fitted by a polynomial of degree 11 (red) for the low magnetic field range ($B^{-1} > 0.45$ T$^{-1}$), where the two oscillatory components $F_1$ and $F_2$ are superimposed. **b,** Experimental result of $G_{xx}$ (black) and the background component fitted by a cubic polynomial (red) for the high magnetic field range ($B^{-1} < 0.2$ T$^{-1}$), where the oscillatory component $F_3$ dominates.

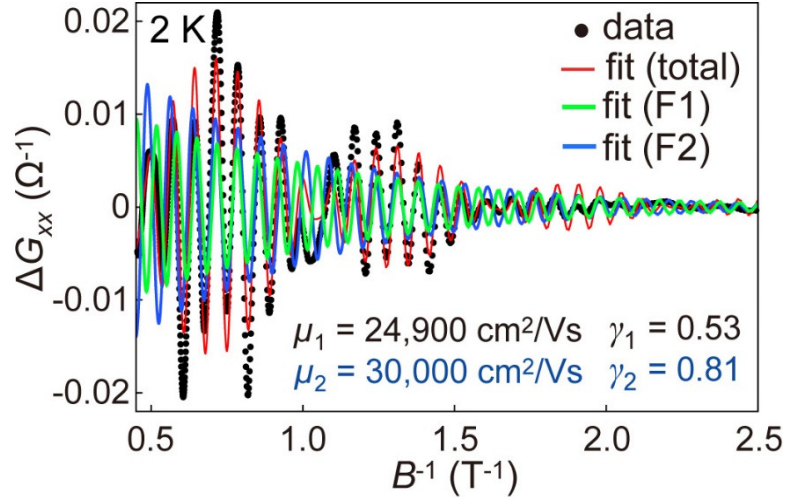

**Supplementary Fig. S4| Fitting to the oscillatory part of conductance $\Delta G_{xx}$ of the 56 ML-thick α-Sn sample in the low magnetic field range ($B^{-1} > 0.45$ T$^{-1}$) by the standard LK theory.** In the $\Delta G_{xx} - B^{-1}$ data at $T = 2$ K (black circles) in the low magnetic field range ($B^{-1} > 0.45$ T$^{-1}$), where only the components $F_1$ and $F_2$ are superimposed, we perform fitting by a sum (red curve) of two oscillations expressed by eq (1) with $G_{01}, \tau_{D1}, \gamma_1$, and $G_{02}, \tau_{D2}, \gamma_2$ as the fitting parameters (green curve and blue curve, respectively). The cyclotron mass values were obtained separately and kept fixed in this fitting.

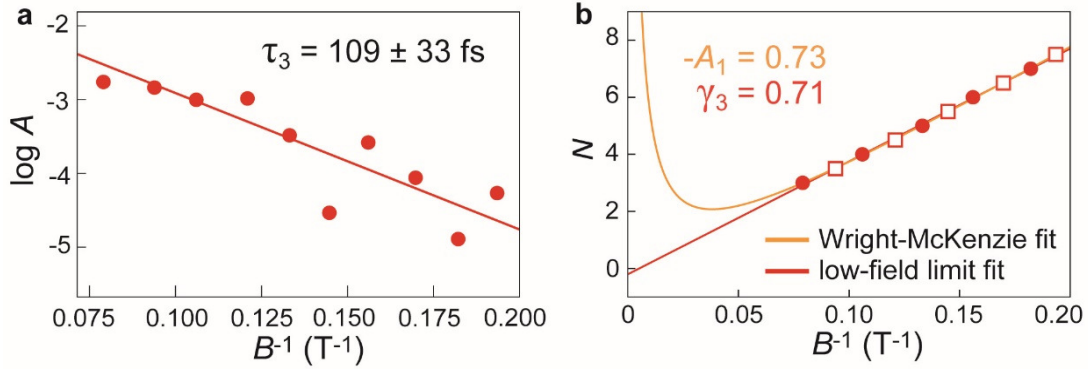

**Supplementary Fig. S5| Analysis of the component $F_3$ in the 56 ML-thick α-Sn sample. a,** Dingle plot of log $A$ vs. $B^{-1}$, where $A$ is given in eq. (8) as $\frac{\Delta G_{xx}}{\frac{2\pi^2 k_B T\hbar\omega_C}{\sinh 2\pi^2 k_B T\hbar\omega_C}}$. Fitting eq (8) (red line) to the experimental data (red circles) gives an estimation of the quantum relaxation time $\tau_3$ = 109 fs, from which we estimate the quantum mobility $\mu_3$ = 1700 cm$^2$/Vs. **b,** Fan plot of component $F_3$. The data points (red circles and white squares, which are taken at the minima and maxima of $\Delta G_{xx}$ in Fig. 2g, respectively) are fitted by two different models given in eq. (9) (red line) and eq. (10) (Wright-McKenzie fit, yellow curve). The Wright-McKenzie fit gives the phase shift $-A_1$ = 0.73, which is in good agreement with that ($\gamma_3$ = 0.71) obtained by eq. (9).

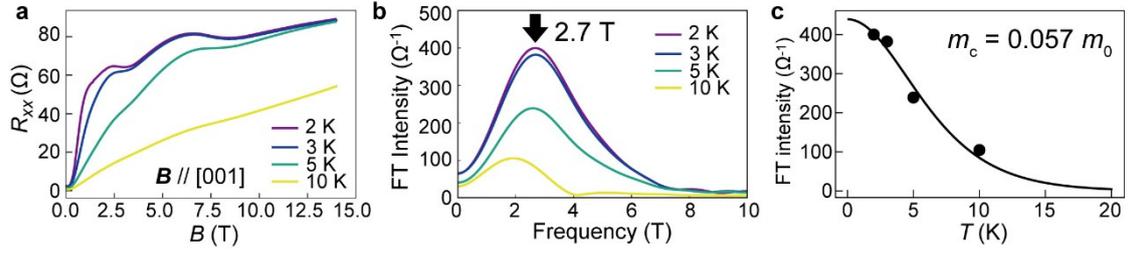

**Supplementary Fig. S6| SdH oscillation in the 56 ML-thick α-Sn under an in-plane magnetic field. a,** Temperature dependence of the longitudinal resistance $R_{xx}$ in α-Sn under an in-plane magnetic field along $[\bar{1}10]$ (***B**//**I***). **b,** Fourier transformed (FT) spectra of $\Delta G_{xx}$ at various temperatures. **c,** Peak intensities of the FT spectra *vs.* temperature fitted by the standard LK theory (Methods).

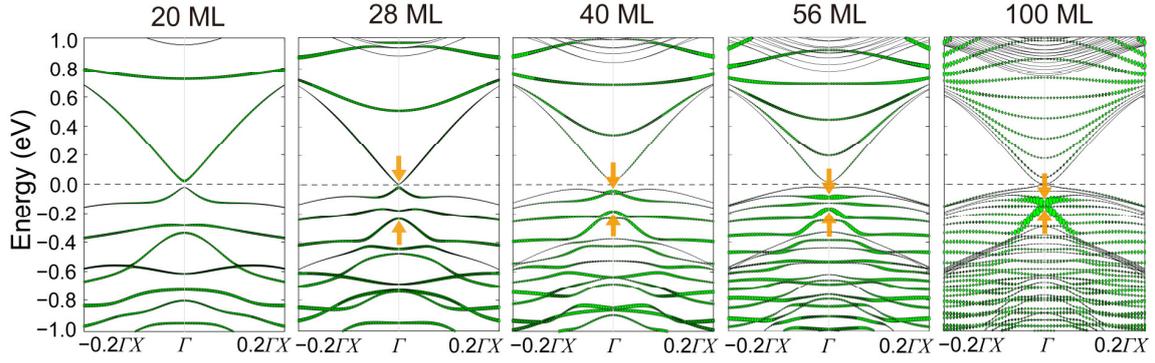

**Supplementary Fig. S7| Band structure of α-Sn thin films with various thicknesses.** These are obtained by first-principles calculation. The green points are the contribution of the topmost Sn monolayer (topological surface states). The size of the green points shows the magnitude of the contribution of the surface state. A gap (pointed by orange arrows) appears in the TSS with decreasing the thickness, which is possibly induced by the hybridization of the top and bottom surface states, and partly by the large compressive strain (–0.76%) in the samples, as shown in Supplementary Fig. S7.

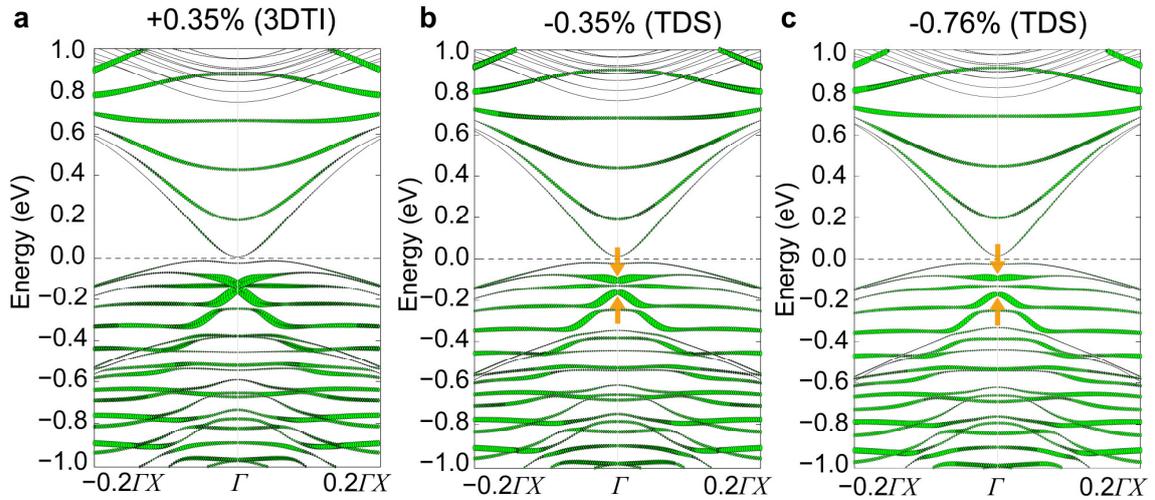

**Supplementary Fig. S8 | Band structure of 56 ML-thick α-Sn films under different in-plane strain: a, 0.35% (tensile strain, topological insulator (TI)). b, –0.35% (compressive strain, topological Dirac semimetal (TDS)). c, –0.76% (compressive strain, TDS).** These are obtained by first-principles calculation. The green points are the contribution of the topmost Sn monolayer (topological surface states). The size of the green points shows the magnitude of the contribution of the surface state. There is a gap opening (pointed by orange arrows), which is larger with a larger compressive strain. These results indicate that the large compressive strain (–0.76%) in the samples is partly responsible for the gap opening in the TSS.

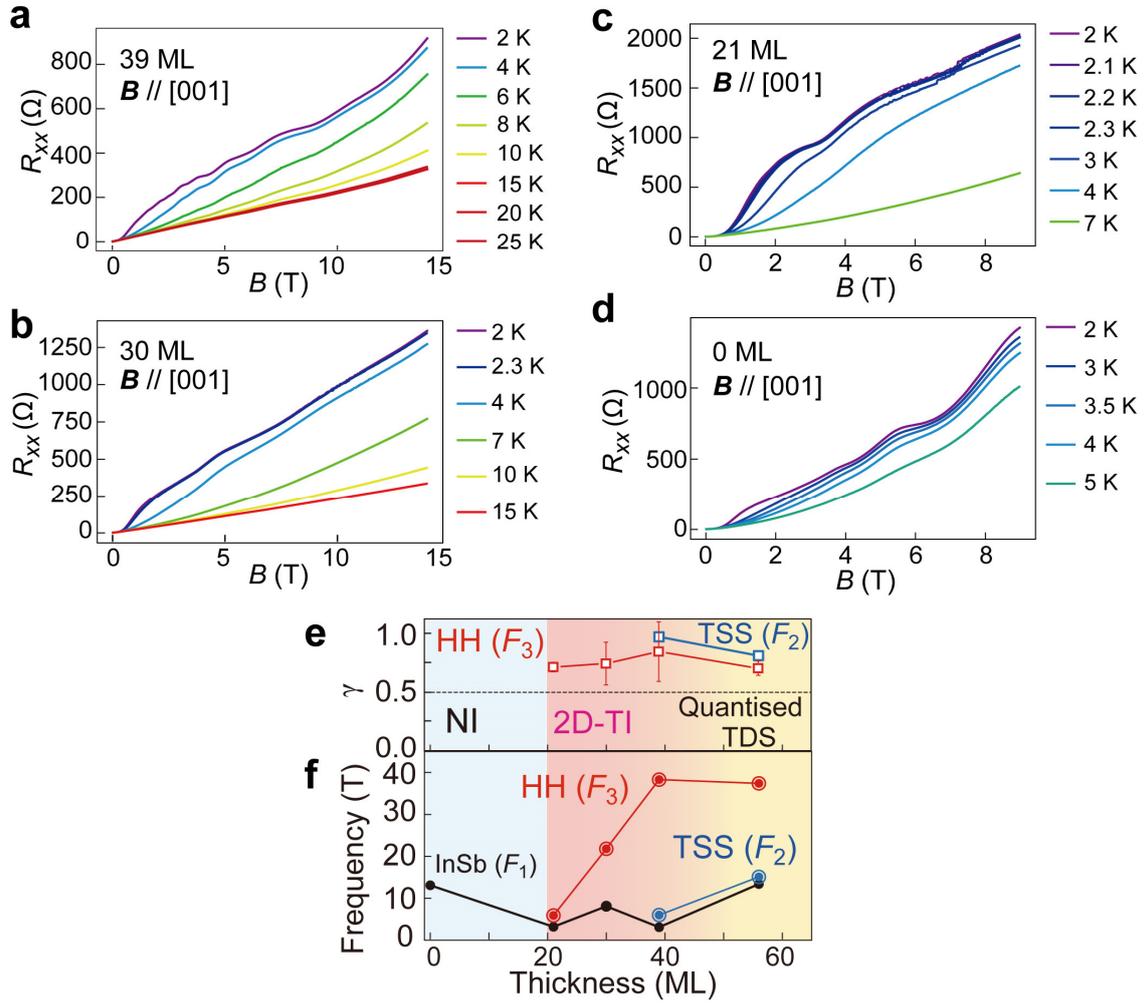

**Supplementary Fig. S9| Thickness dependence of the SdH oscillations. a-d,** SdH oscillations at various temperatures in α-Sn thin films with various thicknesses (39, 30, 21, 0 ML). **e,f,** Thickness dependence of the frequencies and phase shift values $\gamma$ deduced from the SdH oscillations, which is already shown in Fig. 4c. The phase shift of the HH band remains non-trivial values within the error bar in all the thicknesses.

## Supplementary Note 1: Gap opening in the topological surface state (TSS)

The phase shift $\gamma_2$ estimated from the SdH oscillation of the TSS is 0.8098, which is deviated from the ideal value of 1. Our calculated results suggest that this deviation may be caused by a gap opening in the TSS. Supplementary Fig. S7 shows calculated results of the surface state (green points) of α-Sn, whose complete bulk and surface structures are shown in Fig. 4a. A gap (pointed by orange arrows in Fig. S7) appears in the TSS with decreasing the thickness, which suggests that the gap is possibly induced by the hybridization of the top and bottom surface states. Furthermore, our calculations shown in Supplementary Fig. S8 indicate that the gap in TSS is enhanced when we increase the in-plane compressive strain in α-Sn. These results thus suggest that the large compressive strain (–0.76%) in our α-Sn samples is partly responsible for the gap opening in the TSS.

## Supplementary Note 2: Angular dependence of component $F_1$ (InSb)

Angular dependence of the frequency $F_1$ seems to vary proportionally to $1/\cos\theta$ up to $\theta = 60°$, as shown in Fig. 3b. However, under an in-plane magnetic field ($\theta = 90°$), there remains one oscillation (Supplementary Figs. S6). The fitting results of this in-plane $\Delta G_{xx}$ by the Lifshitz-Kosevich (LK) theory indicate a frequency $F_{\text{inplane}} = 2.7$ T and a cyclotron mass $m_{\text{c,inplane}} = 0.057\ m_0$. We attribute this in-plane component $F_{\text{inplane}}$ to a parallel conduction in the bulk InSb substrate. This complicated angular dependence may reflect a coexistence of several origins of the parallel conduction in both the InSb substrate and the InSb buffer. We note that the conduction channel in the InSb buffer may be a two-dimensional carrier gas, which is formed at the InSb side of the α-Sn/InSb interface due to a possible band bending. These InSb-related parallel conductions are difficult to distinguish. However, the trivial topological phase shift and light cyclotron mass of the $F_1$ component likely indicate that this component originates from the conduction band of InSb, not from α-Sn.